\documentclass[prl,nofootinbib,superscriptaddress]{revtex4}
\usepackage[a4paper,left=1.5cm,right=1.5cm,top=3cm,bottom=3cm]{geometry}
\usepackage[toc]{appendix} 

\usepackage{amssymb,amsmath,amsfonts}
\usepackage[dvipsnames]{xcolor}
\usepackage{graphicx}
\usepackage{longtable}
\usepackage{verbatim}
\usepackage{color}
\usepackage[mathscr]{euscript}
\usepackage{framed}
\usepackage{amsmath}
\usepackage{mdframed}
\usepackage{amsmath}
\usepackage{amsfonts}
\usepackage{mathrsfs}
\usepackage{amssymb}
\usepackage{mathrsfs}  
\newcommand{\arXiv}[2]{\href{http://arxiv.org/pdf/#1}{{\tt [#2/#1]}}}
\newcommand{\arXivold}[1]{\href{http://arxiv.org/pdf/#1}{{\tt [#1]}}}

\usepackage{color}
\usepackage{hyperref}
\hypersetup{colorlinks, citecolor=bluscuro, linkcolor=black, urlcolor=bluscuro}
\definecolor{rossos}{cmyk}{0,1,1,0.55}
\definecolor{bluscuro}{rgb}{0.15, 0.2, .85}
\definecolor{bluchiaro}{cmyk}{1,.3,0.,0.1}

\graphicspath{{./Figures/}}

\def\ltsim{\lower3pt\hbox{$\, \buildrel < \over \sim \, $}}  
\def\gtsim{\lower3pt\hbox{$\, \buildrel > \over \sim \, $}}  
\newcommand{\be}{\begin{equation}}  
\newcommand{\ee}{\end{equation}}  
\def\ga{\mathrel{\raise.3ex\hbox{$>$\kern-.75em\lower1ex\hbox{$\sim$}}}}  
\def\la{\mathrel{\raise.3ex\hbox{$<$\kern-.75em\lower1ex\hbox{$\sim$}}}}  

\makeatletter   
\@addtoreset{equation}{section}   
\makeatother   
   
\numberwithin{equation}{section}

\newcommand{\dd}{{\rm  d}}

\newcommand{\R}{\zeta}

\newcommand{\mail}[1]{\href{mailto:#1}{{#1}}}

\newcommand{\doi}[2]{\href{https://dx.doi.org/#1}{#2}}

\usepackage[dvipsnames]{xcolor}
\usepackage{hyperref}
\hypersetup{colorlinks, citecolor=ForestGreen, linkcolor=BrickRed, urlcolor=NavyBlue}
\makeatletter   
\@addtoreset{equation}{section}   
\makeatother

\usepackage[normalem]{ulem}

\begin{document}  
  
\baselineskip=18pt   
\begin{titlepage}  
\begin{center}  
\vspace{0.5cm}  
  
\Large {\bf The Primordial Black Hole Formation \\ from Single-Field Inflation
is Not Ruled Out}
  
\vspace*{20mm}  
\normalsize  

{\large\bf 
 Antonio ~Riotto\footnote{\mail{antonio.riotto@unige.ch}}  }

\smallskip   
\medskip

\it{\small $^1$ Department of Theoretical Physics and Gravitational Wave Science Center,  \\
			24 quai E. Ansermet, CH-1211 Geneva 4, Switzerland}

  
  \end{center}
  
\vskip0.6in

\centerline{\bf Abstract}
\vskip 0.5cm  
\noindent
A standard scenario to form primordial black holes in the early universe is based on a  phase of ultra-slow-roll in single-field inflation when the amplitude of the short scale modes is enhanced compared to the CMB plateau.   Based on general arguments, we show that  the  loop corrections to the large-scale  linear power spectrum from the short modes are small and conclude  that the scenario is not ruled out.

\vspace*{2mm}   

\end{titlepage} 

\section{I. Introduction}  \label{sec:intro}
\noindent
 Primordial Black Holes (PBHs) and the possibility that they form all (or a fraction) of the dark matter in the universe have become a hot topic   again \cite{PBH1,PBH2,PBH3,kam,rep2,revPBH}  after the detection of  two $\sim 30 M_\odot$ black holes through the gravitational  waves generated during their merging \cite{ligo}. 
 
 A  standard  mechanism to create   PBHs in the early universe  is  through the enhancement  of the curvature perturbation $\zeta$   at small scales   \cite{s1,s2,s3}. Such an enhancement  can occur within single-field models of inflation \cite{lr} if there is an Ultra Slow Roll (USR)  period in which the inflaton potential is extremely flat. For such a mechanism to work,  an enhancement of the  power spectrum of the curvature perturbation must  occur,  from its $\sim 10^{-9}$ value at large Cosmic Microwave Background (CMB) scales to $\sim 10^{-1}$ on small scales. A reasonable question is therefore what is the impact on the small amplitude large-scale  power spectrum of   loops containing large amplitude   short modes. 
Indeed, if loop  corrections from the short modes are larger than the tree-level  contribution to the large-scale power spectrum, the USR mechanism to generate PBHs will be inevitably  ruled out,  as  recently advocated in Ref. \cite{yokoyama}. 

In this short note we argue that the PBH Formation  from single-field inflation
is not ruled out. The short modes may change the large-scale  power spectrum plateau in two ways: 

\begin{enumerate}
\item  they propagate in an unperturbed universe and a long mode may be 
generated by the superposition of two short modes. 
This contribution is Poisson-like and largely suppressed compared to the large-scale  power spectrum; 

\item alternatively,  the short modes propagate in a universe which is perturbed by the presence of the long CMB mode. However, the time dependence of such linear long CMB mode is insignificant (otherwise the all scenario will be ruled  already at the tree-level!) and the short modes experience its presence as a background which can be absorbed by a short momentum rescaling. The corresponding  loop correction involve the correlation between the short mode expectation value and the long wavelength mode which, in turn, is dictated by the corresponding consistency relation \cite{m} and vanishes  in the limit of exact scale invariance. At this stage, the expert reader might be confused by this statement as it is  well-known that the consistency relation is violated in USR \cite{c}. The crucial point is that this is true for those linear fluctuations whose decaying mode becomes larger than the growing (in fact constant) mode during the USR phase. This does not happen for the linear long CMB fluctuation whose decaying mode is always exponentially 
smaller than the constant mode. We will argue therefore that these  loop corrections are as well not dangerous. 

\end{enumerate}
This short note is organised as follows. In Section II we will elaborate about these general arguments and in Section III we will discuss a specific example.

\section{II. General arguments}
\setcounter{section}{2}
\noindent
We suppose that the universe goes through a period of Slow-Roll (SR), followed by a period of USR, followed again by a period of  SR during inflation and focus on the dynamics of the long linear CMB modes, those which have left the Hubble radius much earlier (about 20 e-folds earlier  for PBHs masses of $\sim 30 M_\odot$) than the USR phase. 
%

Including loop corrections to  the long  linear CMB modes amounts  to solving the non-linear equation for a long mode $\R_L$. It  takes schematically the form \cite{p}

\be
\widehat O[\R_L]={\cal S}[\R_S,\R_L],
\ee
where ${\cal S}$ represents a generic sum of operators which, at one-loop,  are quadratic in the short wavelength modes $\R_S$ and $\widehat O$ is the operator acting on the long mode. 
It is important to notice that,  once  a  short mode goes outside the Hubble radius, it becomes classical in the sense that $[\zeta_S,\zeta_S']\simeq 0$ \cite{qc}  or in the sense that the phase space of states is larger, because of the USR phase, than  $H^2/k_S^2$ ($H$ being the Hubble rate during inflation) which is much larger than the quantum contribution equal to  $1/2$. This is true  for both  fast and slow  transitions \cite{c}, so that we  can treat the mode as a classical variable and no difference between a quantum and a classical approach is expected.

Formally, the solution of the dynamics is given by

\be
\R_L=\widehat O^{-1}\left[{\cal S}[\R_S,\R_L]\right].
\ee
The one-loop power spectrum of the long modes will then be  the sum of two pieces, again schematically

\be
\label{fund}
\langle \R_L\R_L\rangle= \langle \widehat O^{-1}\left[{\cal S}[\R_S,\R_L=0]\right]\widehat O^{-1}\left[{\cal S}[\R_S,\R_L=0]\right]\rangle+\langle \widehat O^{-1}\left[{\cal S}[\R_S,\R_L]\right]\R_L\rangle.
\ee
There are indeed two ways the short modes may alter the long mode power spectrum, which we describe in turn.

\subsection{II.A The first contribution}
\noindent
 The first  term in Eq. (\ref{fund}) corresponds to the case in which the source of the short modes evolve in an unperturbed metric, that is it does not depend on the long mode, but still
 the superposition of two short-scale modes can lead to a long scale mode. Once the long mode  is well outside of the Hubble radius, the various Hubble patches spanned within one wavelength of $\R_L$ are uncorrelated. This means that the fluctuations of the short modes quickly average out and are not able to provide the coherent effect that would be necessary to source a time dependence on large-scales. In order for the short mode diagrams to be able to induce a large correction  to the power spectrum of  the long mode, one needs coherence of the short modes on a comoving distance of the order of the inverse of the long mode momentum $\sim k^{-1}_L$. However, since the short modes are enhanced only on a finite range of momenta which are much larger than $k_L$, the corresponding  fluctuations are coherent at most on a scale $\sim k_S^{-1}$, and therefore the source from the short modes  becomes quickly uncorrelated as the short mode  spans several Hubble regions. 

One expects therefore  the first  contribution to the long mode from the short modes to be independent from the momentum $k_L$, {\it i.e.} Poisson distributed,  and to be suppressed by   the inverse of the number of independent Hubble patches in a box of radius $\sim k^{-1}_L$, that is by $(k_L/k_S)^3$. By dimensional arguments one will get 

\begin{eqnarray}
\delta P_\R(k_L)\sim \frac{A^2_S}{k_S^3} &\simeq& \left(\frac{A_S^2}{A_L}\right)\left(\frac{k_L}{k_S}\right)^3\frac{A_L}{k_L^3}\sim 10^{-17} \frac{A_L}{k_L^3}\lll \frac{A_L}{k_L^3},\nonumber\\
P_\R(k_L)&=&\frac{A_L}{k_L^3},
\end{eqnarray}
where $A_L\sim 10^{-9}$ and $A_S\sim 10^{-1}$ are the amplitudes of the long and short mode power spectra respectively and  we have taken  $k_L\sim 10^{-2}$ Mpc$^{-1}$ and $k_S\sim 10^{6}$ Mpc$^{-1}$ for PBHs of mass  about $10\,M_\odot$.

\subsection{II.B The second contribution}
\noindent
We elaborate now about the second term, in Eq. (\ref{fund}).  Let us consider a source which is quadratic in the short modes, ${\cal S}\sim \zeta_S^2$, as in the example we will describe below and that was adopted in Ref. \cite{yokoyama}. The  two-point correlator of the long modes is  of the form 

\be
\label{fund1}
\langle \R_L\R_L\rangle\sim \langle \widehat O^{-1}\left[{\cal S}[\R_S,\R_L]\right]\R_L\rangle\sim \langle\langle\zeta_S\zeta_S\rangle_{\zeta_L}\zeta_L \rangle, 
\ee
that is proportional to the bispectrum in the squeezed limit. The latter  is determined by the consistency relation   (for the USR case, see Refs. \cite{bravo,f}), where primes denote again derivative with respect to the number of e-folds,

\begin{eqnarray}
\label{cr}
\langle\langle\zeta_S\zeta_S\rangle_{\zeta_L}\zeta_L \rangle&\sim& P'_\zeta(k_L)P'_\zeta(k_S)
-\frac{\dd\ln {\cal P}_\zeta(k_S)}{\dd\ln k_S}P_\zeta(k_S)P_\zeta(k_L), \nonumber\\
{\cal P}_\zeta(k)&=&\frac{k^3}{2\pi^2}P_\zeta(k).
\end{eqnarray}
Such  consistency relation can be easily derived using the ADM parametrisation and adopting the $\zeta$-gauge defined by imposing that the spatial metric take the form $h_{ij}=a^2(N)e^{2\zeta_L(N)}\delta_{ij}$ (where $a$ is the scale factor and $N$ the number of e-folds). This gauge choice leaves some large scale residual gauge choice, one can for example shift  the long mode curvature perturbation  by $\zeta_L(N)\rightarrow \zeta_L(N)+\alpha(N)$ (together with a change in the 
shift vector $N_i$) and rescale the spatial coordinates $x^i\rightarrow e^{\alpha(N)}x^i$. These transformations are responsible for the first and the second  pieces of Eq. (\ref{cr}), respectively. Such a residual symmetry leading  to the consistency relation (\ref{cr}) is valid for long modes which can be time dependent, but whose spatial derivatives are negligible.

We will show in the following that  the   linear long CMB mode is practically constant  in time, that is $P'_\zeta(k_L)\lll P_\zeta(k_L)$, even during the USR phase.  The  consistency relation   then reduces to the one of SR \cite{m} 

\begin{eqnarray}
\label{cr1}
\langle\langle\zeta_S\zeta_S\rangle_{\zeta_L}\zeta_L \rangle&\sim& -\frac{\dd\ln {\cal P}_\zeta(k_S)}{\dd\ln k_S}P_\zeta(k_S)P_\zeta(k_L).
\end{eqnarray}
This can be understood as follows. If the linear long CMB mode is constant in time, 
it  acts as a rescaling of the coordinates and we can absorb it by rescaling  the momenta

\be
k_S\rightarrow \widetilde{k}_S=e^{\R_L}k_S,
\ee
 dictating that  no correlation between short and long modes is present if the short mode power spectrum is scale invariant. In such a case, 
in the  expectation value of the source composed by the short modes,   the loop  is integrated over all the short mode momenta and the  rescaling is irrelevant;  no correlation between the short-scale power and the long mode  exists. This simple argument tells that the short modes may influence  the long mode only when their power spectrum is not exactly scale invariant. 

We expect therefore  (up to a  numerical factor dictated by the time evolution)

\begin{eqnarray}
\label{dd}
\delta P_\R(k_L)&\sim&   P_\R(k_L)\int\frac{\dd^3 k_S}{(2\pi)^3}P_\zeta(k_S)\frac{\dd\ln {\cal P}_\zeta(k_S)}{\dd\ln k_S}=
P_\R(k_L)\int \dd \ln k_S \,{\cal P}_\zeta(k_S) \frac{\dd\ln {\cal P}_\zeta(k_S)}{\dd\ln k_S}\nonumber\\
&=&P_\R(k_L)\Delta  {\cal P}_\zeta,
\end{eqnarray}
where the variation of the dimensionless power spectrum $\Delta  {\cal P}_\zeta$ has to be evaluated in the range where it is not scale invariant. Since its  maximum variation is of the order of the power spectrum of the short modes at its peak,  $\Delta  {\cal P}_\zeta\simeq 10^{-1}$, we expect 
such   a correction to be small as well. 

These  considerations are general and are valid beyond one-loop. Another  important point to remark is that neither IR nor UV divergences are expected since the non scale invariance of the power spectrum is expected only in a finite range of momenta. 

%

\subsection{II.C The constancy over time of the long CMB mode}
\noindent
The  argument proposed earlier regarding the second contribution  is  valid  since the linear long CMB mode is basically constant on super-Hubble scales during the USR phase. 
The equation for the  linear curvature perturbation $\R_{k_L}$ as a function of the number of e-folds $N$ to go till the end of inflation 
reads at the linear level 

\be
\label{in}
\R_{k_L}''+\frac{(a^3\epsilon)'}{(a^3\epsilon)}\R_{k_L}'+\frac{k_L^2}{a^2H^2}\R_{k_L}=0,
\ee
where $\epsilon=-H'/H$ is the slow-roll parameter. For long modes  the solution   is given approximately as 

\be
\R_{k_L}(N)\simeq A_{k_L}+B_{k_L}\int_{N}\frac{\dd N'}{a^3(N')\epsilon(N')},
\ee
where $A_{k_L}$ and $B_{k_L}$ are constant in time. During the standard SR phase,  $(a^3\epsilon)\sim a^3$ and the piece proportional to the constant $B_k$ is identified with the decaying mode. However, when $(a^3\epsilon)'/(a^3\epsilon)$ changes sign, for instance during a period of USR when $(a^3\epsilon)\sim a^{-3}$, the decaying mode is indeed growing. One is always free to include arbitrary
contributions from the decaying mode in the growing
mode. Nonetheless, it is convenient to identify the constant
$A_{k_L}$  as an approximate solution for the
growing mode on sufficiently large-scales. Rewriting Eq. (\ref{in}) in an iterative form,  the lowest
order solution for the  growing mode solution after the USR phase, when there are still $N_f$ e-folds to go till the end of inflation  is given
by \cite{leach}

\be
\R_{k_L}(N_f)\simeq \alpha_{k_L}\R_{k_L}(N_k),
\ee
where

\be
 \alpha_{k_L}\simeq1+ \int_{N_f}^{N_{k_L}}\frac{{\rm d}N'}{a(N')}\frac{a^2(N_{k_L})}{a^2(N')}\frac{\epsilon(N_{k_L})}{\epsilon(N')}.
\ee
and  $N_{k_L}$ is the  time at which the mode $k_L$ leaves the Hubble radius. Only modes which have $|\alpha_{k_L}|\gg 1$ grows during the USR phase. 

Suppose now there is a period of SR, followed by a period of USR, followed in turn by another period of SR. We are interested in modes which exit the Hubble radius approximately 60 e-folds to go till the end of inflation, that is on CMB scales. Typically, the USR period when PBHs are generated takes place at about $N_i\simeq 40$ e-folds (for PBH masses of the order of $10\,M_\odot$) to go and last
for a number of e-folds around unity. We take these figures as indicative, our results will not change for different assumptions. We get

\be
 \alpha_{k_L}-1\simeq e^{-2(N_{k_L}-N_i)}\sqrt{\frac{\epsilon(N_i)}{\epsilon(N_f)}}\simeq e^{-40}\cdot 10^4\simeq e^{-40}\cdot e^{10}=e^{-30}\lll 1.
\ee
%
%
This means that 

\be
\frac{P'_{\zeta}(k_L,N_f)}{P_{\zeta}(k_L,N_f)}\lll 1.
\ee
This simple derivation makes it clear why linear modes which have exited the Hubble radius much earlier than the USR phase, like the ones relative the CMB modes,  suffer no growth   during the USR phase. For them the decreasing mode remain always much smaller than the growing (constant) mode. This  is why the linear  CMB modes do not grow during the 
USR phase, they remain constant in time. Only modes which exit the Hubble radius in the proximity of the USR phase may grow significantly.   Long CMB modes  can  be absorbed by a momentum rescaling of the  short modes.

\section{III. An example}
\setcounter{section}{3}
\noindent
\noindent
 Let us now consider the same  action  as in Ref. \cite{yokoyama}

\be
S[\R]=M_{\rm pl}^2H\int\dd^3x\,\dd N\,a^3\epsilon\left[\R'^2-\frac{(\nabla\R)^2}{(aH)^2}+\frac{1}{2}\eta' \R'\R^2\right],
\ee
where  $\eta=(\epsilon'/\epsilon)$.
The corresponding equation of motion for the long modes in the CMB range  reads


\be
\R_{k_L}''+\frac{(a^3\epsilon)'}{(a^3\epsilon)}\R'_{k_L}
+\frac{1}{4}\frac{(a^3\epsilon\,\eta')'}{(a^3\epsilon)}\int\frac{\dd^3 k_S}{(2\pi)^3}\,\R_{k_S}\R_{\vec{k}_L-\vec{k}_S}=0,
\ee
whose integration leads to

%
%
\begin{eqnarray}
\R_{k_L}&=&-\frac{1}{4}\int^{N}_{N_{i}} \frac{\dd N'}{a^3(N')\epsilon(N')}\int^{N'}_{N_{i }}\dd N''[a^3(N'')\epsilon(N'')\eta'(N'')]'\int\frac{\dd^3 k_S}{(2\pi)^3}\R_{k_S}(N'')\R_{\vec{k}_L-\vec{k}_S}(N'').
\end{eqnarray}
%
%
%
Typical models of single-field inflation which generate  peaks in the power spectrum are characterised by  a rapid transition from the initial SR phase to the subsequent USR like phase, back to the SR phase. During   the USR phase the modes grow approximately as 
$\epsilon^{-1/2}\sim a^3\sim e^{3(N_i-N)}$.  Notice that on super-Hubble scales the time evolution of the parameter $\epsilon$ dictates the time evolution of the short mode fluctuations, even in the case of 
sudden transitions, and  $\zeta_S$ and $\zeta_S'$ commute as operators; they can be considered as classical objects\footnote{As one can check, for instance,  taking the Super-Hubble limit of the mode functions in Eqs. (16) and (17) of Ref. \cite{yokoyama}.}.

The  parameter $\eta$ jumps from small values to $-6$, back to small values  at $N_f$ e-folds to go till the end of inflation, which takes place at $N=0$.
 Under these circumstances, 
we  obtain 

\be
\R_{k_L}(N=0)\simeq -\frac{3}{2}\int\frac{\dd^3 k_S}{(2\pi)^3}\R_{k_S}(N_f)\R_{\vec{k}_L-\vec{k}_S}(N_f).
\ee
%
%
%
 The first type of correction at  one-loop to the power spectrum of the CMB modes comes when the source of the short modes evolve in an unperturbed metric

\be
\delta P_\R(k_L,N=0)=
2\left(\frac{3}{2}\right)^2\int\frac{\dd^3 k_S}{(2\pi)^3}P_{\R}(k_S,N_f)P_{\R}(|\vec{k}_L-\vec{k}_S|,N_f).
\ee
As already argued, in the limit $k_S\gg k_L$, the correction is Poisson-like, that is it does not depend upon $k_L$. This is because the short modes with large amplitude $A_S$ are correlated at most on a scale $\sim k_S^{-1}$. In  the case of a sharp transition  the power spectrum of the short modes may at most have a  growth of ${\cal P}_{\R}(k_S)\sim k^5$ \cite{steep,anat} up to a peak at $k_c$ \cite{steep,anat}. One   then obtains 
what expected from the general arguments in the previous section, that is 

\be
\label{Poisson}
\delta P_\R(k_L,N=0)\sim 
10^{-2}\frac{A_S^2}{k_c^3}=10^{-2}\left(\frac{A_S^2}{A_L}\right)\left(\frac{k_L}{k_c}\right)^3\left(\frac{A_L}{k_L^3}\right)\ll  P_\R(k_L),
\ee
where again we have taken $k_L\sim 10^{-2}$ Mpc$^{-1}$ and $k_c\sim 10^{6}$ Mpc$^{-1}$ for PBH of mass  about $10\,M_\odot$.
Notice also that for the time variation of the long CMB mode of the curvature perturbation  one has
$
\delta P'_{\R}(k_L,N=0)\sim [a^3(N_f)\epsilon(N_f)/a^3(N=0)\epsilon(N=0)]^2\delta P_{\R}(k_L,N=0)\ll P_{\R}(k_L)$
so that in the consistency relation (\ref{cr}) one can comfortably neglect the first term even at higher orders. 
%

The second contribution to the correction to the power spectrum of the long CMB modes comes from correlating the average of the 
source of the short modes in the presence of a long mode $\R_{k_L}$ with another long mode

\be
\langle \R_{k_L}(N=0)\rangle \simeq- \frac{3}{2}\int\frac{\dd^3 \widetilde{k}_S}{(2\pi)^3}\langle\R_{\widetilde{k}_S}(N_f)\R_{-\vec{\widetilde{k}}_S}(N_f)\rangle, \,\,\,\, \widetilde{k}_S=e^{\R_L}k_S.
\ee
which gives the following correction at the end of inflation

\be
\delta P_\R(k_L,N=0)\simeq- \frac{3}{2}P_\R(k_L)\Delta {\cal P}_\zeta(k_S,N_f),
\ee
where  we recall that $\Delta {\cal P}_\zeta(k_S,N_f)$ must be computed over the range where the power spectrum is not scale invariant. Since the fall of the power spectrum
can be at most its initial amplitude, $\Delta {\cal P}_\zeta(k_S,N_f)\sim 10^{-1}$, we infer  that this correction is as well negligible. Notice again that $\delta P'_\R(k_L,N=0)$ is suppressed compared to $\delta P_\R(k_L,N=0)$ by a factor $[a^3(N_f)\epsilon(N_f)/a^3(N=0)\epsilon(N=0)]^2\lll 1$.

One may wonder what happens for the case of a smooth transition. Despite the fact that the parameter $\eta$ and $\eta'$ vary dramatically, the exact solution for the mode functions remain scale invariant up to small corrections ${\cal O}(\eta_V)$ \cite{c}. Indeed, even if the  curvature perturbation still evolves
during the transition, and is finally  fixed after the SR attractor  phase is attained, the resulting power spectrum in this period remains nearly scale invariant.  Taking during the smooth transition $\eta'\simeq 6/(N_f-N_{\rm SR})$ and approximately constant, where  $N_{\rm SR}$ is the time the SR attractor is reached, the integration with time needs to be done from the time of the transition $N_f$ to $N_{\rm SR}$ and we  obtain

\begin{eqnarray}
\delta P_\R(k_L)&\simeq&\frac{\eta_V A_S}{ (N_f-N_{\rm SR})}  P_\R(k_L) \ln\frac{k^{\rm SR}_S}{k^f_S},
\end{eqnarray}
where $k^{\rm SR}_S$ is the short scale corresponding to the moment the SR attractor is attained and $k_S^f$ the scale at which the USR ends. Again this correction is small.

Finally, let us comment on the effect of three extra one-loop contributions which might appear from the cubic interactions and which are not suppressed by gradients on 
super-Hubble scales. The first, proportional to $\zeta_L'\R_S'^2$,  is suppressed by the derivative of the long mode; the second, 
proportional to $\zeta_L\R_S'^2$,  is suppressed by $\epsilon^2$, the third, 
arising from the redefinition $\zeta\rightarrow\zeta_n+\eta\,\zeta_n^2/4+\zeta_n\zeta'_n$ (to remove a boundary term)    \cite{m} is suppressed in the  SR phase following the USR period. The corresponding loops are therefore suppressed.

We conclude that  the PBH formation scenario through a period of USR is not  ruled out.

\vskip 0.5cm
\centerline{\bf Acknowledgements}
\vskip 0.2cm
\noindent
We thank M. Biagetti, R. Bravo, C. Byrnes, V. De Luca  G. Franciolini,  A. Iovino, S. Patil, M. Sasaki, and A. Urbano for useful discussions and the organizers of the  workshop ``Messengers of the very early universe: Gravitational Waves and Primordial Black Holes" (12-14 Dec 2022, 
Centro Universitario Padovano, Padova, Italy) for the nice atmosphere where this work has been completed.
A.R. is supported by the
Boninchi Foundation for the project ``PBHs in the Era of
GW Astronomy".


\end{document}